\pdfoutput=1  
\documentclass[final,3p,times,twocolumn]{elsarticle}
 \biboptions{comma,sort&compress}
\usepackage{here}
 \usepackage{graphicx}
  \usepackage{epsfig}
  \usepackage{hyperref}
\usepackage{amssymb}
\usepackage{amsmath}

\def\beq{\begin{equation}}
\def\eeq{\end{equation}}
\def\bea{\begin{eqnarray}}
\def\eea{\end{eqnarray}}

\journal{Nuc. Phys. (Proc. Suppl.)}

\usepackage{booktabs}
\usepackage{units}
\usepackage{mathabx}	
\usepackage{color} 
\usepackage{xcolor} 
\usepackage{braket}
\usepackage{slashed}

\newcommand{\bls}{\baselineskip}

\newcommand{\dint}{\textrm d}

\newcommand{\dtot}[2]{\frac{\dint {#1}}{\dint {#2}}}
\newcommand{\tdtot}[2]{\tfrac{\dint {#1}}{\dint {#2}}}

\newcommand{\emath}{\,\text{e}}
\newcommand{\Lagrange}{\mathcal L}
\newcommand{\intd}[3]{\int\limits_{#1}^{#2}\!\!\textrm d{#3}\,}  
\newcommand{\intdurch}[2]{\int\!\!\tfrac{\dint{#1}}{#2}\,}
\newcommand{\intdurchl}[4]{\int\limits_{#1}^{#2}\!\!\tfrac{\dint{#3}}{#4}\,}

\newcommand{\Ord}[1]{\mathcal O\!\left({#1}\right)}

\definecolor{Blue}{cmyk}{1,0.5,0,0.22}
\definecolor{Green}{cmyk}{1,0,0.80,0.61}

\newcommand{\hard}[1]{\textcolor{black}{#1}}
\newcommand{\col}[1]{\textcolor{black}{#1}}
\newcommand{\acol}[1]{\textcolor{black}{#1}}
\newcommand{\soft}[1]{\textcolor{black}{#1}}

\begin{document}

\begin{frontmatter}

 \title{Massive Boson Production at Small q$_T$ in Soft-Collinear Effective
 Theory}
 
 \author[label1]{Thomas Becher}
 \author[label2]{Matthias Neubert}
 \author[label2]{Daniel Wilhelm}
 \address[label1]{Institut f\"ur Theoretische Physik, Universit\"at Bern, Switzerland}
 \address[label2]{Institut f\"ur Physik (THEP), Johannes Gutenberg-Universit\"at Mainz, Germany}

\begin{abstract}
\noindent
We study the differential cross sections for electroweak gauge-boson and 
Higgs production at small and very small transverse-momentum $q_T$.
Large logarithms are resummed using soft-collinear effective theory.
The collinear anomaly generates a non-perturbative scale $q_*$, which protects 
the processes from receiving large long-distance hadronic contributions.
A numerical comparison of our predictions with data on the transverse-momentum 
distribution in Z-boson production at the Tevatron and LHC is given.
\end{abstract}

\end{frontmatter}

\section{Drell-Yan-Like Processes}
Historically the Drell-Yan (DY) process \cite{PhysRevLett.25.316} denoted
the inclusive production of a virtual photon by quark-antiquark annihilation in
hadron collisions and the subsequent decay into a lepton pair.
Its main features are
strongly coupled initial and color-neutral final states
and so the photon case can easily be generalized to W- and Z-boson production.
Even the Higgs production via gluon fusion can be described in a similar way.\\
The transverse-momentum distribution of DY-like processes is
one of the most basic observables at hadron colliders. It is used e.g. to extract
the W-boson mass and width and is of great phenomenological relevance
for Higgs-production at the LHC. Especially the regime of small
transverse-momentum $q_T^2\ll M^2$ is important, because it gives the largest
contribution to the total cross section. Here $q_T$ denotes the transverse component of the boson 4-momentum $q$, while $M^2$
is its invariant mass $q^2$. We thus consider:
\[\dtot{\sigma}{q_T}\quad\text{with}\quad
q^2=\hard{M^2}\gg 
\col{q_T^2}\gg\Lambda_{QCD}^2\,.\]\\[-1\bls]
The hierarchy in this regime between the hard scale $\hard{M}$ and the
collinear scale $\col{q_T}$ leads to large logarithms which spoil the
perturbativity of fixed-order calculations. These logarithms need to be
resummed to all orders in perturbation theory to achieve a predictive result.\\
Our approach \cite{1109.6027-Becher-2011} is to
factorize the cross section using an effective field theory (EFT) and resum
large logarithms via renormalization group (RG) techniques. The
appropriate EFT to describe DY-like processes is the soft-collinear effective
theory (SCET)~\cite{Bauer:2000yr}, because it accounts for the complex
structure of underlying scales originating from Sudakov double logarithms \cite{Sudakov:1954sw}.
\section{Factorization using SCET} 
SCET is an EFT of QCD. In general it describes any number of collinear
modes, high energetic particles (or Jets) with light-like momenta and soft
modes, which mediate the only interactions between the different collinear
fields.\vspace{0.2\baselineskip}
\begin{figure}[htb]
\flushleft 
\includegraphics[width=\linewidth]{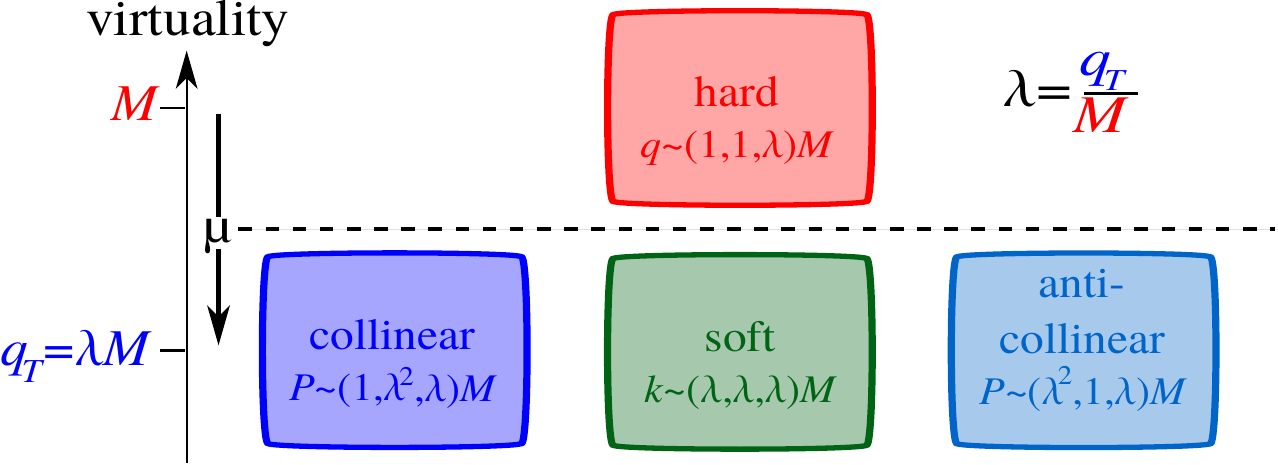}\\[-0.9\bls]
  \caption{Momentum modes in SCET.}
  \label{020-SCETII}  
\end{figure}  

In DY-like processes there are two collinear modes, defined by the two opposite
light-like momenta of the colliding hadrons. The different
momentum regions are best defined in lightcone coordinates. Therefore we
introduce two light-like reference vectors $\col{n}$ and $\acol{\bar n}$ along
the beam axis with $\col{n}\cdot\acol{\bar n}=2$. Now every 4-vector $k$ can be
decomposed into its collinear ($\col{k_+}$), anti-collinear ($\acol{k_-}$) and perpendicular ($k_\perp$) component, by
projecting it onto $\col{n}$ and $\acol{\bar n}$.

The values of interest are the
virtuality $\sqrt{k^2}$ and the scalings of momenta:\[\text{Scaling:} 
\quad k\sim\left(\col{k_+},\acol{k_-},k_T\right)
\quad\text{ with }  
\quad k_T^2=-k_\perp^2\,. 
\] A virtuality of $\Ord{M}$ identifies the hard modes, which are integrated
out like the produced DY-boson.
The scaling is used to distinguish between the different collinear and soft
modes (Figure \ref{020-SCETII}).

Up to power suppressed terms, the factorization using SCET leads to the
following double differential cross section, where $y$ denotes the rapidity of the DY-boson:
\[
\dtot{^2\sigma}{ q_T\dint y} \sim\hard{H}
\cdot \sum_{ij}Q_{ij}
\cdot \intd{}{}{^2\vec x_\perp}
\emath^{-i\vec q_\perp\vec x_\perp}
\cdot \soft{W}\cdot
\col{\mathcal B_{i/N_1}}
\acol{\mathcal B_{j/N_2}}\,.
\]
It consists of a hard function $\hard{H}$,
a sum over contributing partons and effective charges,
a soft function $\soft{W}$
and two collinear functions $\col{\mathcal B}$.
The hard function contains the Wilson coefficients of the EFT.
The soft function leads to scaleless integrals,
thus does not contribute to all orders in perturbation theory:
\begin{figure}[t]
\flushleft
  \includegraphics[width=\linewidth]{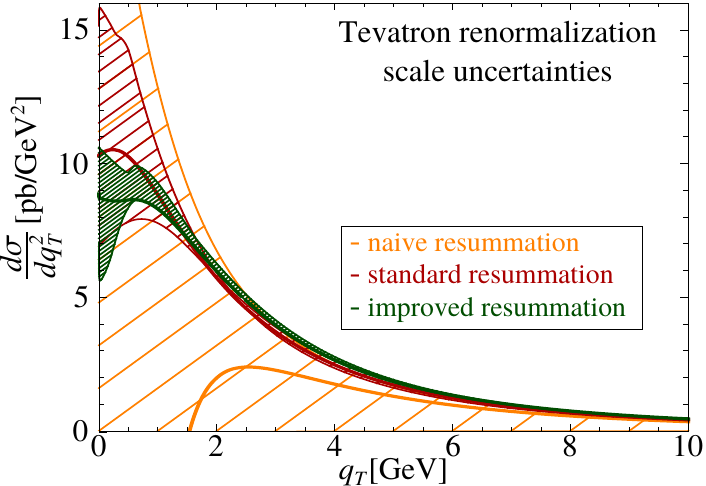}\\[-0.9\bls]
  \caption{Scale uncertainties for different resummation schemes.}
  \label{plResSchemes}
\end{figure}   
\begin{align*}
\hard{H(M,\mu)=\left|C(-M^2,\mu^2)\right|^2}\,,
&&\soft{W=1+\Ord{\lambda^2}}\,.
\end{align*}
Comparing the collinear functions $\col{\mathcal B}$ with the representation of
ordinary parton distribution functions (PDF) in SCET, it turns out they are just
generalized $x_T$ dependent PDFs (gPDF):
\vspace{-0.5em}\begin{align*}
\col{\mathcal B_{q/N}(\xi,L_\perp)=} 
&\col{\intdurch{t}{2\pi}\emath^{-i tnp} 
\bra{N\!\:}
\bar\chi_c(nt+\textcolor{black}{x_\perp})
\tfrac{\slashed{\bar n}}2
\chi_c(0)
\ket{N}}\,.
\end{align*}
Here the $x_T$ and $\mu$ dependence is hidden in the logarithm
$\col{L_\perp}=\ln\left(\col{x_T^2}\mu^2\right)$. The Wilson coefficients are
known, the soft corrections vanish and one can match the gPDFs on partonic
level onto ordinary PDFs, only missing long-distance hadronic effects of
$\Ord{\Lambda_{NP}^2x_T^2}$:
\[
\col{\mathcal B_{i/N}(\xi,L_\perp)=
\sum_{j}\intdurchl{\xi}{1}{z}{z}}\col{\mathcal I_{ij}(z,L_\perp)
\phi_{j/N}(\tfrac{\xi}z,\mu)}\,.
\]
\section{Collinear Anomaly and Resummation}
\begin{figure}[t]
\flushleft
  \includegraphics[width=\linewidth]{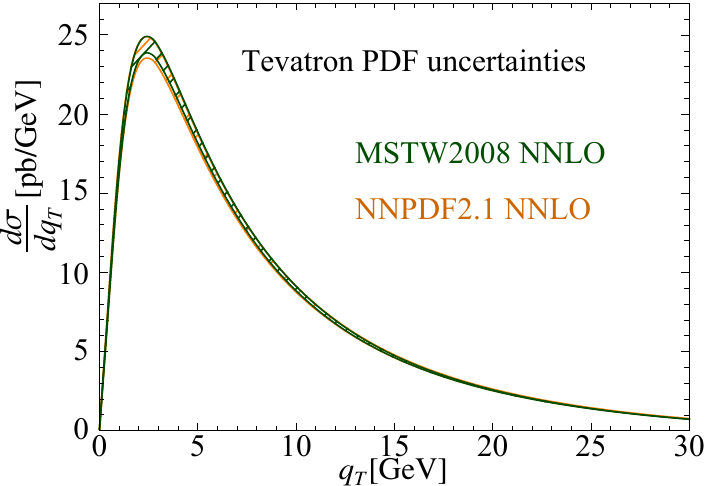}\\[-0.9\bls]
  \caption{PDF uncertainties for different PDF sets.}
  \label{plPDF}
\end{figure}  
On the classical level the SCET Lagrangian respects the so-called rescaling
symmetry. Since the two collinear fields can not interact with each other,
each of their Lagrangians is invariant under the rescaling of momenta of the
other one. At higher orders the collinear anomaly (CA) appears, the
symmetry is broken by quantum corrections and restricted to joint rescaling,
which introduces an unexpected invariant:
\begin{align*}
\begin{array}{lc}
\col{\Lagrange_{\text{c}}}:&
\acol{\bar p}\;\to\;\bar \alpha\acol{\bar p}\\
\acol{\Lagrange_{\bar{\text{c}}}}:&
\col{p}\;\to\;\alpha\col{p}
\end{array}
\;\xrightarrow{CA}\; \alpha\cdot\bar\alpha\equiv 1
\quad\Rightarrow\quad
\hard{M^2}=2\col{p}\acol{\bar p}\,.
\end{align*}
It turns out that this directly effects the matching of the gPDFs by generating
a power-like dependence on the hard scale $\hard{M}$:
\vspace{-0.5em}\[
\col{\mathcal B_{i/N}}\acol{\mathcal B_{j/N}}
\;\xrightarrow{CA}\;
\hard{(x_T^2M^2)^{F_{ij}(L_\perp)}}
\col{B_{i/N}(\xi,L_\perp)}\acol{B_{j/N}(\xi,L_\perp)}\,.
\]\\[-1.2\bls]  
This term ensures the RG invariance in the absence of soft
contributions and  is important for the resummation of large logarithms.

The resummation of the hard function is simply done by using the RG equation:
\[
\hard{H(M},\mu\hard{)}=\hard{H(M,\mu_h)}\cdot
U(\hard{\mu_h},\mu)
\quad\text{ and set }\quad\hard{\mu_h\sim M}\,.
\]
Resumming the terms under the Fourier integral,
\[\intd{}{}{x_T^2}
\emath^{-iq_Tx_T}
\hard{(x_T^2M^2)^{F_{ij}(L_\perp)}}
\col{B_{i/N}(\xi,L_\perp)}\acol{B_{j/N}(\xi,L_\perp)}\,,
\] 
is more subtle.
It contains two types of logarithms,
$\col{L_\perp}=\ln\left(\col{x_T^2}\mu^2\right)$ from the collinear modes and
$
\ln\tfrac{\hard{M^2}}{\mu^2}$ from the CA. Setting
$\mu\sim\col{q_T}$ and expanding in $\alpha_s$ leads to small $\col{L_\perp}$,
because $x_T$ is the conjugate variable of $q_T$ under the Fourier integral.
At small $\col{q_T}$ this naive resummation scheme leads to a large logarithm
contained in $\hard\eta\sim\alpha_s\ln\tfrac{\hard{M^2}}{\mu^2}$, which spoils the perturbativity.\\
To avoid this, our standard resummation scheme is to count $\hard{\eta}$ as
$\Ord{1}$ and include higher order terms (in $\alpha_s$) of the CA.\\
The standard resummation breaks down when $\hard{\eta}$ reaches~1. This
happens at the scale $\col{q_*}$:
\[
\col{q_*^Z\approx \unit[1.8]{GeV}}\,,
\hspace{2em}
\col{q_*^H\approx \unit[7.7]{GeV}}\,.
\]
To lower $\col{q_T}$ beyond $q_*$ one has to dismiss the demand of small
$\col{L_\perp}$ by setting $\mu\sim q_*$. The appearing higher order terms of
the CA form a Gaussian under the integral, which regulates it independently of
$\col{q_T}$, even at vanishing transverse-momentum
$q_*\gg\Lambda_{QCD}>\col{q_T}\geq0$. 
\begin{figure}[t!]
\flushleft
  \includegraphics[width=\linewidth]{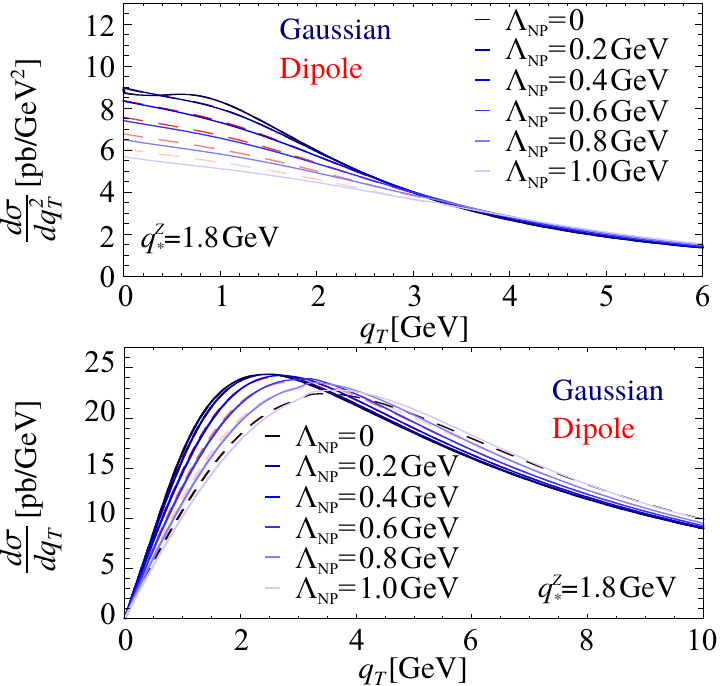}\\[-0.9\bls]
  \caption{Impact of hadronic effects on the intercept (top) and the
  peak-region (bottom) for Z-production.}
  \label{plNP}
\end{figure}  

\section{Uncertainties}
The first plot (Figure\ref{plResSchemes}) shows the renormalization scale
uncertainties for Z-boson production at the Tevatron. The error bands correspond
to varying the default renormalization scale $\mu_d=\col{q_T+q_*^Z}$ by a factor of two. The
 errors of the naive resummation (orange) lead to unpredictive results, because
 the terms of the CA are missing. The standard resummation (red) gives considerable
 smaller error bands. The errors of the improved resummation (green) are
 somewhat smaller above $\col{q_*^Z}$, but significantly below.
 As a consequence all following plots are made using the improved resummation
 scheme.
  
 The second plot (Figure \ref{plPDF}) shows the PDF uncertainties for two
 different PDF sets at the Tevatron. The different shape of the plot 
 depends on showing $\tdtot{\sigma}{q_T}$ instead of $\tdtot{\sigma}{q_T^2}$. The first is used
 to point out the peak region, the latter for the intercept. The error bands
 correspond to one standard deviation to the center value.
 The uncertainties are around 5\%, therefore lie within the renormalization scale
 uncertainties.
 \begin{figure}[t!]  
\flushleft  
  \includegraphics[width=\linewidth]{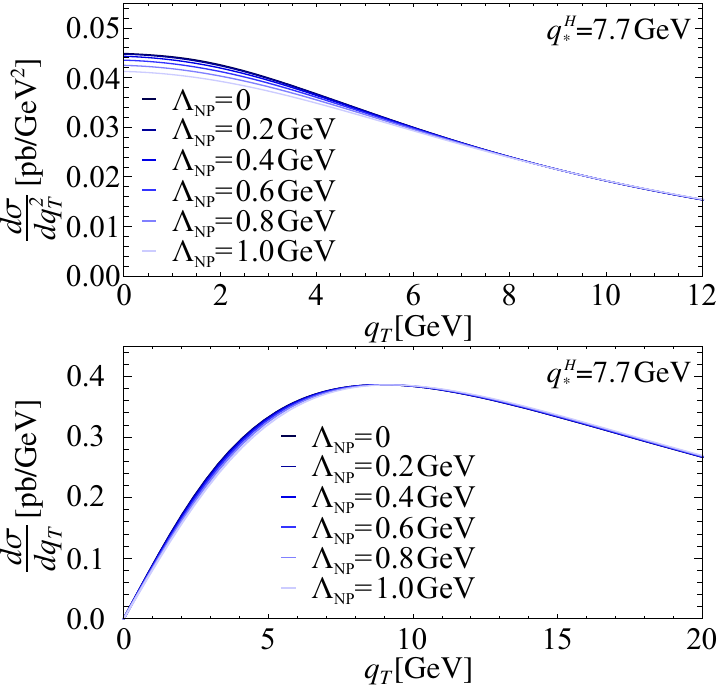}\\[-0.9\bls]
  \caption{Impact of hadronic effects on the intercept (top) and the
  peak-region (bottom) for Higgs-production.}
  \label{plNPHiggs}
\end{figure}   

 The hard function is independent of $\col{q_T}$, thus can be regarded as an
 overall factor with constant uncertainties (Table \ref{tabhard}). In the
 following plots the error bands correspond to the scale uncertainty.
     \begin{table}[h]
\begin{center}
\vspace{-0.5\bls} 
\begin{tabular}{lcc}
 & $\mu_h^2=m_Z^2$ &  $\mu_h^2=-m_Z^2$  \\ \hline
NLL & $1.000^{+0.160}_{-0.060}$ & $1.334^{+0.201}_{-0.074}$ \\ 
NNLL & $1.087^{+0.010}_{-0.001}$ & $1.131^{+0.001}_{-0.014}$ \\ 
N$^3$LL & $1.119^{+0.006}_{-0.001}$ & $1.130^{+0.001}_{-0.001}$
\end{tabular}
\end{center}
\vspace{-0.9\bls}\caption{\label{tabhard} 
The hard function $H\left(M_Z,\mu\right)$ at $\mu=M_Z$ for space-like and
time-like choices of $\mu_h^2$. The uncertainties are obtained by varying
$\mu_h$ by a factor two about the default value.}
\vspace{-1\bls}
\end{table} 

\section{Long-Distance Hadronic Effects} 
We model the non-perturbative effects with a Gaussian (blue) and a Dipole
(red) factor in the gPDFs. The plots in Figure \ref{plNP} show the impact of
these effects on the intercept (top) and the peak region (bottom). By adjusting
$\Lambda_{NP}$ we can fit the peak region onto experimental data without
influencing the measurable rest of the cross section (the intercept can not be measured at
hadron colliders). Since these effects should be universal, we can set
$\Lambda_{NP}$ in one measurement and use it as input for other distributions.
All following plots are made using the Gaussian model, because the differences
between the two models are marginal.

 Figure \ref{plNPHiggs} shows the same plots in the Higgs case.
 The effects are strongly suppressed compared to the Z-boson production in Figure
 \ref{plNP}. The DY-like cross sections are protected from receiving large
 long-distance hadronic contributions by the CA. The effects scale with the ratio $\nicefrac{\Lambda_{NP}}{q_*}$, which is much smaller in Higgs production. 

\section{Final Results}
\begin{figure}[t]
\flushleft
  \includegraphics[width=\linewidth]{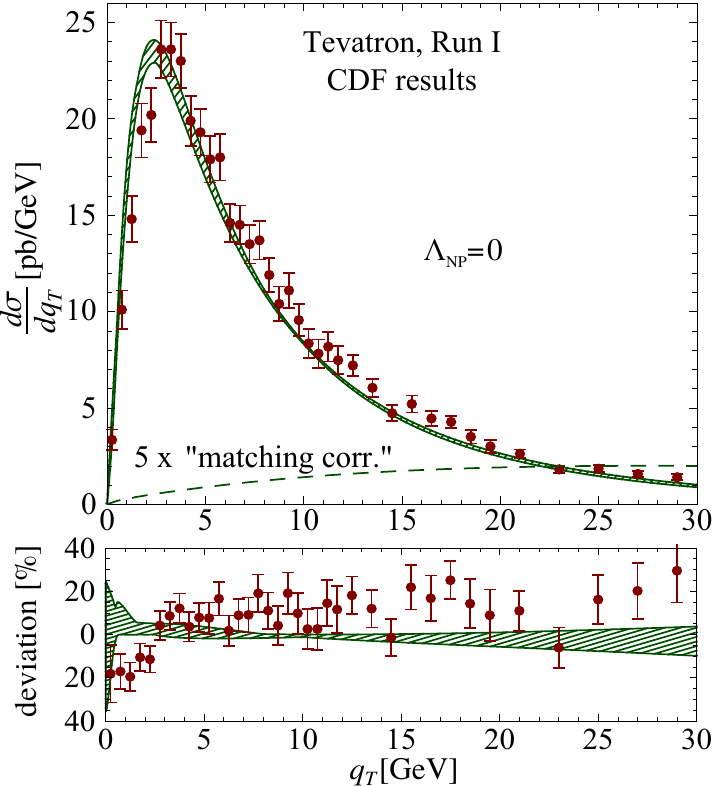}\\[-0.9\bls]
  \caption{Comparison with data on Z-boson transverse-momentum distribution at
  CDF \cite{Affolder:1999jh} without hadronic effects.}
  \label{plTeva1}
\end{figure}   
\begin{figure}[t]
\flushleft
  \includegraphics[width=\linewidth]{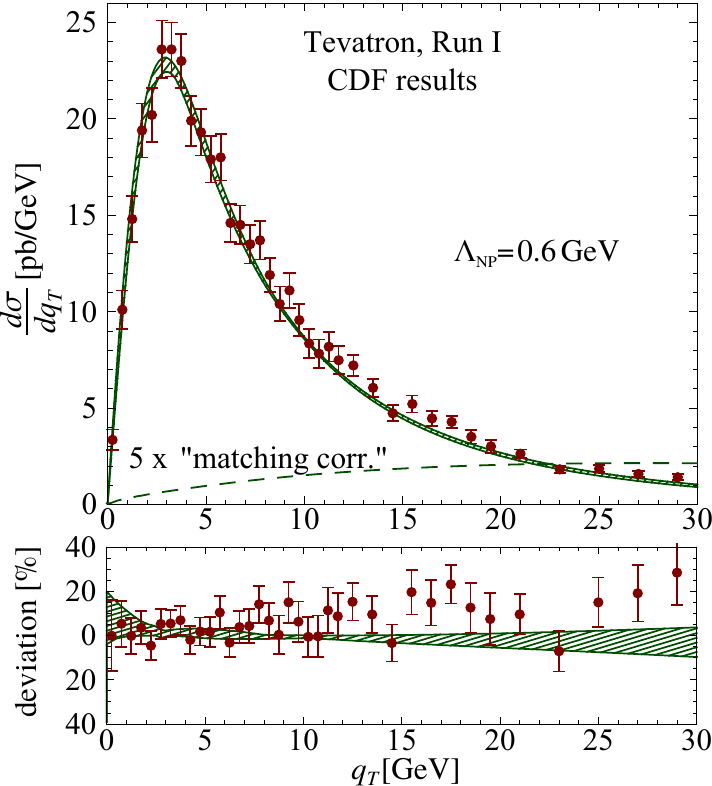}\\[-0.9\bls]
  \caption{Comparison with data on Z-boson transverse-momentum distribution at
  CDF \cite{Affolder:1999jh} with hadronic effects.}
  \label{plTeva2} 
\end{figure}     
The last two plots show our final results, comparisons with data on Z-boson
transverse-momentum distribution at CDF \cite{Affolder:1999jh}, Figure
\ref{plTeva1} without and Figure \ref{plTeva2} with hadronic effects. Including
these effects obviously improves the agreement of theory and data at small
$\col{q_T}$, while it does not influence the predictions above
$\col{q_T}\approx\unit[15]{GeV}$. By using SCET we miss terms of
$\Ord{\lambda^2}$, which become important at large $q_T$. To receive a result
for the whole $q_T$-region, we match our result onto fixed-order calculations.
The deviation at larger $q_T$ arises because we only include matching at NLO
fixed-order and should be reduced at NNLO.
The matching correction is shown five times
larger to make it visible.
 
\section{Conclusion}
As shown in the last plots, our approach of factorizing the DY-like cross
sections, using SCET and resumming large logarithms via RG-methods, leads to
very good agreement of theory predictions and experimental data, together with
small scale uncertainties. There have been a lot of approaches since the first
resummation \cite{NuclPhysB.250.199} in 1985, but this is the first time it was
done directly in momentum space and it is free of Landau-pole singularities. Two
important advantages of this approach are, it is straightforward to extend the calculation
to higher orders in $\alpha_s$ and $\lambda$ and the used methods are process
independent, therefore applicable to other problems \cite{Becher:2012qa}.

\bibliographystyle{DISproc}
\bibliography{wilhelm_daniel}

\end{document}